\documentclass[fleqn,twoside]{article}
\usepackage{espcrc2}

\usepackage{graphicx}
\usepackage[figuresright]{rotating}

\newcommand{\AmS}{{\protect\the\textfont2
  A\kern-.1667em\lower.5ex\hbox{M}\kern-.125emS}}

\title{Double Chooz: Optimizing CHOOZ for a possible $\theta_{13}$ measurement}

\author{S. A. Dazeley, for the Double Chooz Collaboration \\
\addressmark[MCSD]{202 Nicholson Hall,\\
    Department of Physics and Astronomy,\\ 
        Louisiana State University, \\ 
        Baton Rouge, Louisiana, 70803, USA}}

\begin{document}

\begin{abstract}
The proposed Double Chooz $\theta_{13}$ experiment is described.
Double Chooz will be an optimized reactor disappearance 
experiment similar to the original CHOOZ.  
The optimization includes an increase in the signal to noise
by increasing the target
volume to twice the original CHOOZ, reducing singles background
with a non-scintillating oil 
buffer region around the target
and carefully controlling
systematic uncertainties 
by measuring the $\bar{\nu_{e}}$ 
flux of the source with a near detector.   The Double Chooz far detector
will be situated in the same cavern as 
CHOOZ but will detect $\sim 50000$ $\bar{\nu_{e}}$s in three years
of operation.  We estimate a systematic uncertainty of
$0.6\%$, and a reduction of the upper limit on $sin^{2} (2 \theta_{13})$
to $0.03$. 
\vspace{1pc}
\end{abstract}

\maketitle

\section{INTRODUCTION}

In recent years, Super-Kamiokande, SNO and KamLAND
have shown that the three known types of neutrino 
($\nu_{e}, \nu_{\mu}, \nu_{\tau}$) have mass and undergo flavor oscillation.  
In particular, $\nu_{e} \rightarrow \nu_{\mu}$ or $\nu_{\tau}$ oscillations
were proven by SNO \cite{SNO}, KamLAND \cite{kamland} 
measured the oscillation
parameters
$\Delta m_{12}^{2} = 7.9^{+0.6}_{-0.5} \times 10^{-5}$eV$^{2}$
and $sin^{2} 2\theta_{12} = 0.82 \pm 0.07$, and
Super-Kamiokande measured the 
\mbox{$\nu_{\mu} \rightarrow \nu_{\tau}$}
oscillation parameters
($1.5 \times 10^{-3} < \Delta m_{23}^{2} < 3.4 \times 10^{-3}$eV$^{2}$ and
$sin^{2} 2\theta_{23} > 0.92$ \cite{SK_K2K}).
Other interesting issues remain.
Strangely,
both $\theta_{12}$ and $\theta_{23}$ are close to maximal
but $\theta_{13}$ is constrained to be small by CHOOZ.
Additionally, neutrinos may violate CP conservation.  
In experiments that search for $\nu_{e}$ appearance in a $\nu_{\mu}$
beam
the CP violating term and $\theta_{13}$ are degenerate.
Disappearance experiments that measure the $\bar{\nu_{e}}$ remaining 
in a $\bar{\nu_{e}}$ source are sensitive only to $\theta_{13}$,
allowing a clean measurement.
In the following, ``CHOOZ'' will refer to the original 
experiment \cite{CHOOZ}, ``Double Chooz'' 
to the new proposal.

\section{IMPROVEMENTS TO CHOOZ}

The most important improvement to CHOOZ will be a near detector
situated $\sim 150$ meters
from the reactor cores.
Its inner detector will be 
identical to that of the far detector enabling a precise
measurement of the unoscillated $\bar{\nu_{e}}$ signal.
Its outer detector will incorporate an extra veto covering
the top and sides to account for the higher muon flux
at its shallow depth ($\sim 65$ meters water equivalent).

CHOOZ has the best upper limit on $\theta_{13}$ of
\mbox{$sin^{2} 2\theta_{13} < 0.2$} at \mbox{
$\Delta m^{2}_{13} = 2 \times 10^{-3}$eV$^{2}$}.
The $\bar{\nu_{e}}$s were detected via
the inverse beta decay reaction  $\nu_{e} + p \rightarrow e^{+} + n$,
giving a prompt signal due to the $e^{+}$ and a delayed
signal $\sim 30 \mu s$ later from neutron capture on
Gadolinium (Gd).  Gd doped scintillator has a
short capture time and increased neutron capture
detection efficiency due to the large amount of energy
released ($\sim 8$ MeV).
Unfortunately, nitrates used to dissolve the Gd caused a reaction with
the scintillator that colored it, limiting the lifetime of 
the experiment.  
Since the CHOOZ result,
R\&D work for LENS at the Max Plank Institute has developed
two classes of Gd loaded scintillator suitable for a new generation of
reactor neutrino experiment.

The signal to noise ratio of Double Chooz will be $\sim100$,
four times better than CHOOZ.  This results from doubling the 
target volume and providing a non-scintillating buffer zone between the
target and the PMTs to reduce background.  

\section{SYSTEMATIC UNCERTAINTIES AND BACKGROUNDS}


Table~\ref{table:1} shows the systematic uncertainties expected.  
The differences
between the near and far
detectors must be small in order to achieve 
our goal of 0.6\% total relative 
systematic uncertainty.  
Particular care must be
taken to ensure that both detectors have the same number of target
protons.  Therefore both target vessels will be made to the same specifications
by the same vendor
at the same time.  They will then be installed, fully assembled, into
their respective detectors and filled from the same batch of scintillator.
The scintillator will be carefully weighed during the filling process.

The improved detector design and lower background rate
will enable a simplification of the analysis relative to CHOOZ.
The Double Chooz
analysis will require only
energy and time cuts.

\begin{table}[htb]
\caption{Summary of systematic uncertainties
 at \mbox{Double Chooz} }
\label{table:1}
\newcommand{\m}{\hphantom{$-$}}
\newcommand{\cc}[1]{\multicolumn{1}{c}{#1}}

\renewcommand{\tabcolsep}{2pc} 
\begin{tabular}{@{}ll}
\hline
          & Goal\\
\hline
Solid Angle                & \m0.2\% \\
Number of Protons       & \m0.3\% \\
Neutron Efficiency   & \m0.2\% \\
Neutron Energy Cut   & \m0.2\% \\
Time Cut   & \m0.1\% \\
Dead Time   & \m0.2\%  \\
Acquisition   & \m0.1\% \\
Background    & \m0.2\%      \\
\hline

Total         & \m0.6\%         \\

\hline
\end{tabular}
\end{table}

\vspace{-9pt}


A singles event rate at the level of 1Hz will be achieved
with U, Th and K concentrations of
$\sim 10^{-12}$, $\sim 10^{-12}$ and $\sim 10^{-10}$ grams/gram
respectively in the scintillator and 
$\sim 10^{-10}$, $\sim 10^{-10}$ and $\sim 10^{-8}$ grams/gram
respectively in the acrylic vessels.
Background estimates
from the rock were scaled from the CHOOZ experiment while those of the PMTs
were obtained from simulations which included Hamamatsu estimates.

The correlated backgrounds can also be estimated by scaling from the CHOOZ
experiment using reactor off data.  The $^{9}Li$ and $^{8}He$ rate measured 
in CHOOZ was \mbox{$\sim 0.2 /$ day}.  Doubling the target volume gives 
\mbox{$\sim 0.4 /$ day} 
in Double Chooz.  

\section{SENSITIVITY LIMIT}

\begin{figure}[htb]
\framebox[75mm]{\rule[-1mm]{0mm}{43mm}\includegraphics[width=17pc,height=14pc]{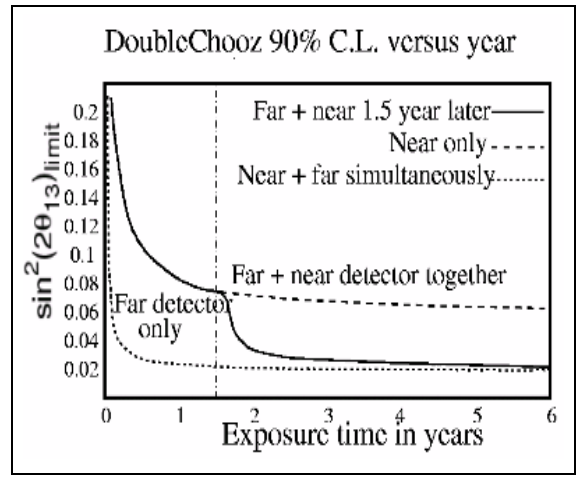}}
\caption{The $sin^{2} (2\theta_{13}) $
sensitivity limit of Double Chooz assuming the real value of
$\theta_{13}$ is zero.}
\label{figure:1}
\end{figure}

The sensitivity limit of Double Chooz is shown in Figure~\ref{figure:1},
assuming an achievable systematic uncertainty of ~0.6\% and a $\bar{\nu_{e}}$
signal rate of $\sim 50,000$ events over three years.  
The old CHOOZ limit is surpassed
within a few months, even before the near detector is built.  The sensitivity
after three years with the near detector will be $sin^{2} 2 \theta \sim 0.03$,
giving an $\sim 85\%$ reduction in the CHOOZ limit and 
excellent discovery potential..

\end{document}